\documentclass[twocolumn]{revtex4}
\usepackage{amssymb}
\usepackage{amsmath}
\usepackage{graphicx}
\usepackage{color}

\begin{document}

\title{Stochastic thermostats: comparison of local and global schemes}
\author{Giovanni Bussi}
\email{gbussi@ethz.ch}                                 
\author{Michele Parrinello}                            
\affiliation{Computational Science, Department of Chemistry and Applied Biosciences,                                                   
ETH Z\"urich, USI Campus, Via Giuseppe Buffi 13, CH-6900 Lugano, Switzerland}
                                                       
\begin{abstract}                                       
We show that a recently introduced stochastic thermostat
[J.~Chem.~Phys.~{\bf 126}, 014101 (2007)] can be considered as
a global version of the Langevin thermostat.           
We compare the global scheme and the local one (Langevin)
from a formal point of view and through practical calculations
on a model Lennard-Jones liquid.                       
At variance with the local scheme, the global thermostat preserves
the dynamical properties for a wide range of coupling parameters,
and allows for a faster sampling of the phase-space.
\end{abstract}

\maketitle

The most common approaches to isothermal molecular dynamics
are perhaps those based on the introduction of an extended
Lagrangian. The root of all these schemes is the Nos\'e{} algorithm
\cite{nose84jcp}, often used in the Hoover formulation \cite{hoov85pra}.
This scheme can be rigorously shown to provide the correct Boltzmann
distribution and has a conserved quantity,
which can be used to check the integration timestep.
A major drawback of the Nos\'e-Hoover method is that it is
not ergodic in some difficult cases, such as harmonic systems.
Several different extensions of the Nos\'e-Hoover method
have been introduced, the most notable one being the
so-called Nos\'e-Hoover chains \cite{mart92jcp},
which addresses the ergodicity issue at the price of an increased complexity
in the algorithm.
Although the Nos\'e-Hoover scheme was originally written as a global thermostat,
i.e. coupled only to the
total kinetic energy of the system, it is sometimes implemented
in a local manner (also called massive Nos\'e-Hoover),
i.e. using an independent thermostat on each degree of freedom
\cite{tobi+93jpc}.

Another common choice is the weak-coupling method, introduced by
Berendsen \emph{et al} \cite{bere+84jcp}. This scheme is a continuous version of
the velocity-rescaling scheme, thus it is a global thermostat.
It is deterministic, stable and easy to implement,
but it does not produce configurations in the
canonical ensemble.

An alternative approach to canonical sampling
is to use stochastic molecular dynamics.
The most common form is Langevin dynamics \cite{schn-stol78prb}.
The Langevin thermostat is local, and
its major feature is that ergodicity can be
proven also in pathological cases.
However, since the friction and noise terms alter significantly the
Hamiltonian dynamics, it
cannot be used to compute dynamical properties, unless an extremely
small friction is used.
Moreover, the effect of the friction and noise terms on the
sampling efficiency is non trivial. Even in applications where dynamical
properties are not relevant it can be difficult to properly tune
the friction in order to achieve an efficient sampling.

In a recent paper~\cite{buss+07jcp} we proposed a stochastic velocity
rescaling which can be considered as Berendsen thermostat plus a stochastic
correction leading to canonical sampling. We also
showed that, in spite of its stochastic nature,
one can define a conserved quantity.
This scheme does not suffer of ergodicity problems in solids~\cite{buss+07jcp},
has been used in practical applications
for equilibration purposes~\cite{dona-gall07prl} or to perform
ensemble averages~\cite{brun+07jpcb,bard+08prl}
and can be combined with variable-cell dynamics to perform
simulations in the isothermal-isobaric ensemble~\cite{zyko+08jcp}.
In the present paper we present an alternative derivation of the same scheme,
where stochastic velocity rescaling is obtained starting from Langevin
dynamics and minimizing the disturbance of the thermostat on the
Hamiltonian trajectory, nevertheless retaining the same thermalization
speed of Langevin dynamics.
This idea was also used by Berendsen \emph{et al} to derive their
algorithm \cite{bere+84jcp}.
Moreover, we show how stochastic
velocity rescaling can be considered as a global version of Langevin
dynamics. Thus the relationship between the two schemes is similar to that
between standard Nos\'e-Hoover and massive Nos\'e-Hoover.
Finally, we compare in practical
situations the efficiency of the local (Langevin) and global (rescaling)
versions, and show that the disruption of Hamiltonian dynamics
observed using Langevin thermostat
is not due to the the stochastic nature of the algorithm but to the use
of a local thermostat.

\section{Continuous equations of motion}

We consider a system described by coordinates $q_i$ and momenta $p_i$,
where $i$ runs over the $N_f$ degrees of freedom,
and with $q$ and $p$ we indicate the set of coordinates $q_i$ and $p_i$.
We associate a mass $m_i$ to each degree of freedom, and
we define a Hamiltonian $H(p,q)=K(p)+U(q)$,
where $U(q)$ is the potential energy,
and $K(p)=\sum_i \frac{p^2_i}{2m_i}$ is the kinetic energy.
We want to sample the canonical distribution
$P(p,q)dpdq\propto e^{-\beta (K(p)+U(q))}$,
where $\beta$ is the inverse temperature,
by means of equations of motion in the form
\begin{subequations}
\label{eq-langevin}
\begin{align}
dp_i(t) & = -\frac{\partial U}{\partial q_i}dt + g_i(t)dt \\
dq_i(t) & = \frac{p_i(t)}{m_i}dt.
\end{align}
\end{subequations}
Equations~\eqref{eq-langevin}
are Hamilton equations
plus a correction force $g_i(t)$ which artificially modifies the
dynamics of the system.
Since the total energy $H$ is conserved in Hamiltonian dynamics,
only $g_i(t)$ is responsible for its variations
and leads to the system thermalization.

In standard Langevin dynamics, the correction force is
\begin{equation}
\label{eq-langevin-force}
g_i(t)dt=- \gamma p_i(t)dt + \sqrt{\frac{2m_i\gamma}{\beta}}dW_i(t),
\end{equation}
where $\gamma$ is the friction coefficient,
and $dW_i(t)$ is a vector of $N_f$ independent Wiener noises,
normalized as $\langle \frac{dW_i(t)}{dt} \frac{dW_j(t')}{dt}\rangle = \delta(t-t')\delta_{ij}$.
The thermalization speed can be quantified calculating the
time derivative of the total energy from Eqs.~\eqref{eq-langevin}
and~\eqref{eq-langevin-force}.
Using the Itoh chain rule \cite{gard03book} we obtain
\begin{multline}
\label{eq-dh}
dH(t)=
\sum_i\left(-\frac{\gamma p_i^2(t)}{m_i}dt
+ \sqrt{\frac{2\gamma }{\beta}\frac{p_i^2(t)}{m_i}}dW_i(t)\right) \\
+ \frac{\gamma N_f}{\beta}dt.
\end{multline}
This expression can be further simplified defining
the average kinetic energy $\bar{K}=N_f(2\beta)^{-1}$,
a relaxation time $\tau=(2\gamma)^{-1}$,
and exploiting
the fact that the noise terms on different degrees of freedom are independent
of each other:
\begin{equation}
\label{eq-dh2}
dH(t)=-\frac{K(t)-\bar{K}}{\tau}dt + 2\sqrt{\frac{\bar{K}K(t)}{N_f\tau}} dW(t).
\end{equation}
It is worth noting that while in Eqs.~\eqref{eq-langevin} and \eqref{eq-dh} there are
$N_f$ independent noise terms, in Eq.~\eqref{eq-dh2} a single noise term is present.

\begin{figure}
\begin{center}
\includegraphics[clip,width=0.3\textwidth]{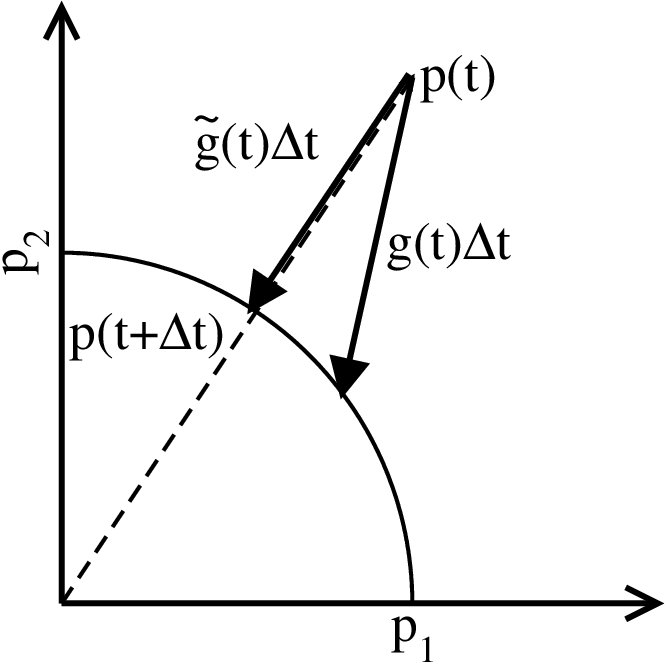}
\end{center}
\caption{
\label{fig-scheme}
Schematic representation of the momentum components at time $t$
and at time $t+\Delta t$. $g(t)$ is a generic force applied to the system.
$\tilde{g}(t)$ is a force which leads to the same change in the kinetic
energy as $g(t)$, but minimizes the disturbance.
}
\end{figure}
We now want to design a new correction force $\tilde{g}_i(t)$ which gives
the same variation of the total energy as Langevin dynamics,
thus the same thermalization speed,
but minimizes the disturbance on the trajectory.
This procedure is exactly the same used by Berendsen \emph{et al}
\cite{bere+84jcp}, the only difference
being that Eq.~\eqref{eq-dh2} on the total energy now
contains a stochastic term.
We first notice that, since the force only acts on the momenta,
fixing a value for $H$ is equivalent to fixing a value for $K$.
Following Ref.~\cite{bere+84jcp} we quantify
the disturbance as $\sum_i m_i^{-1}(\tilde{g}_i(t)dt)^2$.
As it is seen in Fig.~\ref{fig-scheme},
the minimal disturbance for a fixed kinetic energy increment
is obtained
with a force $\tilde{g}_i(t)$ which is proportional to $p_i(t)$.
Thus $\tilde{g}_i(t)=\lambda(t) p_i(t)$, where $\lambda(t)$
is chosen so as to enforce a given variation of the total energy.
Since $\lambda(t)$ includes
a stochastic part, and since the variation of the total energy depends
on the momenta only through the kinetic energy $K$,
this last relation can be written as
\begin{equation}
\label{eq-g-global}
\tilde{g}_i(t)dt=p_i(t)\left[A(K(t))dt+B(K(t))dW(t)\right],
\end{equation}
where $A(K)$ and $B(K)$ are arbitrary functions of the kinetic energy.
The change in the total energy is then
\begin{equation}
\label{eq-dh3}
dH=2A(K)K dt+2B(K)KdW+B^2(K)Kdt.
\end{equation}
Expressions for $A(K)$ and $B(K)$ can be obtained setting
Eq.~\eqref{eq-dh2} equal to Eq.~\eqref{eq-dh3},
resulting in the correction force
\begin{multline}
\label{eq-global-force}
\tilde{g}_idt=\frac{1}{2\tau}\left[
\left(1-\frac{1}{N_f}\right)\frac{\bar{K}}{K}
-1
\right]p_i dt \\
+\sqrt{\frac{\bar{K}}{N_fK\tau}}p_idW.
\end{multline}
This equation is stochastic, with the same noise term used on all
the particles. It is also very similar to the expression of the
force in the Berendsen algorithm.

The combination of Eqs.~\eqref{eq-langevin} and~\eqref{eq-global-force}
results in a continuous, stochastic dynamics which can be shown to
sample exactly the canonical ensemble.
The effect of a $\tilde{g}_i$ parallel to $p_i$
is the same of a rescaling procedure and
the enforced increment of the total energy
in Eq.~\eqref{eq-dh2} is the same as in Ref.~\cite{buss+07jcp}.
Thus, Eqs.~\eqref{eq-langevin} and~\eqref{eq-global-force}
represents the continuous version
of the velocity rescaling described in Ref.~\cite{buss+07jcp}.

Notably, if $N_f=1$, Eq.~\eqref{eq-global-force} becomes equivalent to
Eq.~\eqref{eq-langevin-force}. Thus when
the thermostat is applied to a single degree of freedom,
it is completely equivalent to a Langevin thermostat.
One can perform Langevin molecular dynamics by applying
a thermostat per degree of freedom, or stochastic rescaling
by applying a single thermostat to the total kinetic energy.
Intermediate schemes can be designed, where a thermostat is applied
on each molecule or group of atoms.

\section{Finite timestep algorithm}
In the practical implementation,
time is incremented in discrete steps, and
the Trotter decomposition scheme
\cite{tuck+92jcp,buss-parr07pre}
can be used to separate the integration of Hamilton equations and
the update of the momenta due to $\tilde{g}$.
The former is then integrated using standard velocity-Verlet,
while for the latter we need to integrate
Eq.~\eqref{eq-global-force}.
A possible approach is to consider the propagation of
kinetic energy when the momenta evolve according
to Eq.~\eqref{eq-global-force}, as it is done
in the appendix of Ref.~\cite{buss+07jcp}.
The analytical solution of Eq.~\eqref{eq-global-force}
for a finite time $\Delta t$ is
\begin{subequations}
\label{eq-propagator}
\begin{equation}
p_i(t+\Delta t)
=
\alpha(t) p_i(t),
\end{equation}
where
\begin{multline}
\alpha^2(t) = 
c+\frac{(1-c)(S_{N_f-1}(t)+R^2(t))\bar{K}}{N_fK(t)} \\
+ 2R(t)\sqrt{\frac{c(1-c)\bar{K}}{N_fK(t)}}.
\end{multline}
\end{subequations}
Here $c=e^{-2\gamma \Delta t}=e^{-\Delta t /\tau}$, $R(t)$ is a Gaussian number
with unitary variance and $S_{N_f-1}$ is the sum of $N_f-1$
independent, squared, Gaussian numbers.
Equation~\eqref{eq-propagator}
has been obtained enforcing the evolution of the kinetic energy, thus,
strictly speaking, it does not fix the sign of $\alpha$.
A more rigorous analysis shows that the sign of $\alpha$ should be chosen as
\begin{equation}
\label{eq-alpha-sign}
\text{sign} (\alpha(t))=\text{sign}\left(R(t)+\sqrt{\frac{cN_fK(t)}{(1-c)\bar{K}}}\right),
\end{equation}
to keep into account the finite probability to observe a flip
of the momenta $p_i$ when the force in Eq.~\eqref{eq-global-force} is applied.
The Gaussian number in Eq.~\eqref{eq-alpha-sign} needs to be the same that
is used in Eq.~\eqref{eq-propagator}.
The probability to observe the flip is extremely small
 if $N_f$ is large
and $c\approx 1$, which is the usual case when the thermostat is used as global
and $\tau>\Delta t$.
This is the case in Ref.~\cite{buss+07jcp}, where we set $\alpha>0$.
On the other hand, when the thermostat acts on a few
degrees of freedom,
the sign of $\alpha$ needs to be
calculated by means of Eq.~\eqref{eq-alpha-sign}.
This is always the case for Langevin dynamics.
With simple manipulation,
it can be shown that for $N_f=1$ the integration scheme
given by Eqs.~\eqref{eq-propagator}
and~\eqref{eq-alpha-sign}, combined with velocity-Verlet,
is completely equivalent to the integration
scheme for Langevin dynamics introduced in Ref.~\cite{buss-parr07pre}.

\section{Examples}

Up to now we simply established a theoretical relationship between Langevin dynamics
and stochastic rescaling. The outcome is that the effect of the two algorithms on the
total energy should be equivalent if the friction in the Langevin dynamics
and the relaxation time in the stochastic scale are related by
$\tau=(2\gamma)^{-1}$. However, the stochastic rescaling is expected to give better
dynamical properties, since only the component of the force which changes
the total energy is retained. We test here this affirmations on a simple test-case,
namely a Lennard-Jones fluid with density $\rho=.8442$ and temperature $\beta^{-1}=0.722$,
which is close to the triple point.
Throughout this section we use
reduced Lennard-Jones units for temperature, distance and time.
We simulate a box containing 108 particles,
with periodic boundary conditions, and we cut the interaction at
distance 2.5. We set the timestep to $\Delta t=0.005$, which leads to
a reasonable conservation of the effective energy~\cite{buss+07jcp,buss-parr07pre}.
We then perform runs of $10^{7}$ steps, with both the local
scheme (Langevin dynamics) and the global one (stochastic rescaling),
using a broad range of values of the thermostat
relaxation time $\tau$.

\begin{figure}
\includegraphics[clip,width=0.45\textwidth]{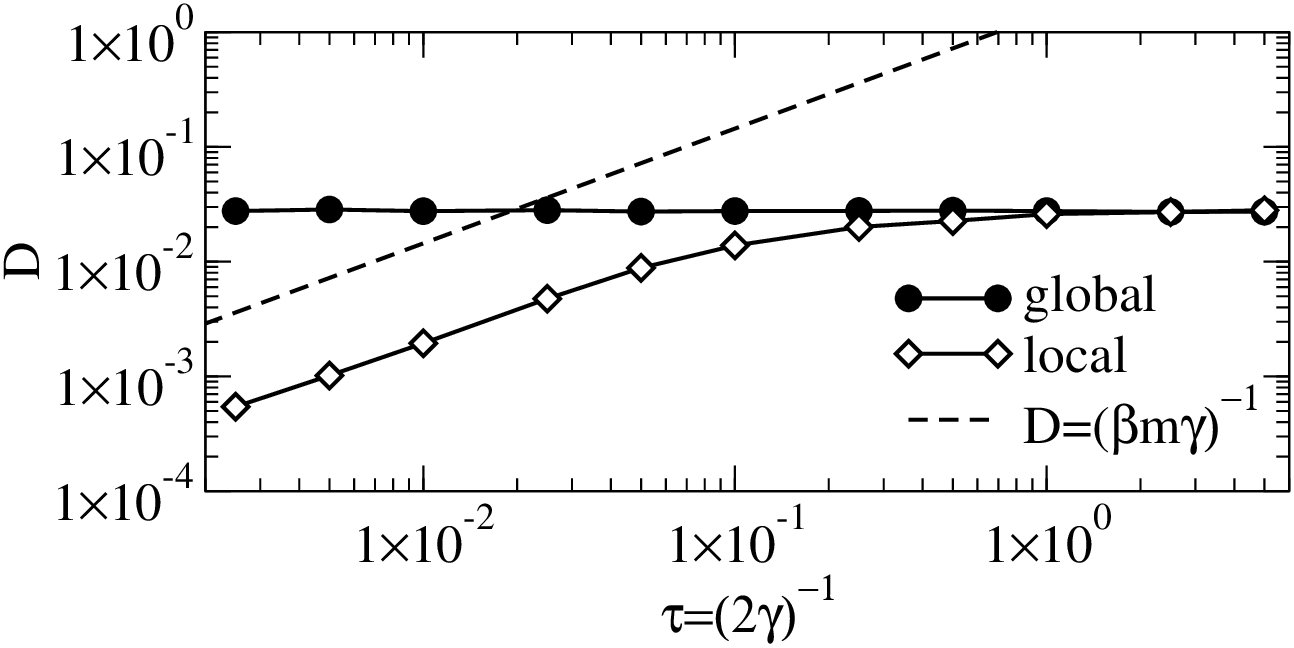}
\caption{
\label{fig-diffusion}
Diffusion coefficient as obtained from thermostated simulations
as a function of the thermostat relaxation time $\tau$, 
for the local and global thermostats as indicated. The statistical error
is smaller than the symbol size.
As a reference, the diffusion coefficient of a free particle
subject to the same Langevin equation is plotted (dashed line).
All the quantities are in Lennard-Jones reduced units.
}
\end{figure}
To quantify the disturbance on Hamiltonian dynamics,
in Fig.~\ref{fig-diffusion} we show
the diffusion coefficient as a function of $\tau$, as
obtained from the Einstein relations.
When the local thermostat is used with a short
relaxation time, the diffusion is strongly quenched.
This happens when the typical collision
time with the external bath $\gamma^{-1}$ is shorter than the
typical collision time between the particles, so that the former
becomes the real bottleneck for the diffusion process.
In the limit of short $\tau$, the equations
of motion tend to a high-friction Langevin dynamics.
In this case, the observed diffusion coefficient
$D$ is proportional to $(\beta m \gamma)^{-1}$,
which is the value of the diffusion coefficient for a free particle
subject to the same Langevin dynamics.
The prefactor is related to the difficulty in crossing the barriers between
different liquid configurations.
On the contrary,
with the global thermostat $D$ is almost independent on $\tau$,
indicating that the disturbance on the dynamics is very small and that
good estimates of $D$ in the canonical ensemble
can be obtained also with a thermostated simulation.

To quantify the equilibration speed we calculate the autocorrelation
time of a few global observables.
The efficiency of a sampling algorithm is optimal
when the autocorrelation time $\tau_X$ of the desired quantity
$X$ is minimal.
If $X$ is the total energy,
$\tau_X$ also indicates how fast
a simulation started from
an unlikely configuration is equilibrated.
We integrate the autocorrelation function
using a windowing function
\begin{equation}
\label{eq-autocorrelation}
\tau_X=
\int_0^Tdt \frac{\langle \delta X(0) \delta X(t) \rangle}{\langle \delta X(0) \delta X(0) \rangle}\left(1-\frac{t}{T}\right),
\end{equation}
where $\delta X=X-\langle X \rangle$ and $T$ is a large value.
The windowing function in parethesis helps the convergence
because it weights less the points with larger statistical error.
Moreover, Equation~\eqref{eq-autocorrelation}
is exactly equal to $T\epsilon^2(T)/(2\langle\delta X^2\rangle)$,
where $\epsilon^2(T)$ is the mean square error from a time
average of length $T$.
The relative accuracy in evaluation of $\tau_X$ is approximately
$\sqrt{2T/T'}$, where $T'=5\times10^4$ is the total run length.
In the following we choose $T=50$, and we expect a relative
accuracy on $\tau_X$ on the order of 5\%.
Since $T\gg\tau_X$, $\tau_X$ is also a good approximation for
the autocorrelation time.

\begin{figure}
\includegraphics[clip,width=0.45\textwidth]{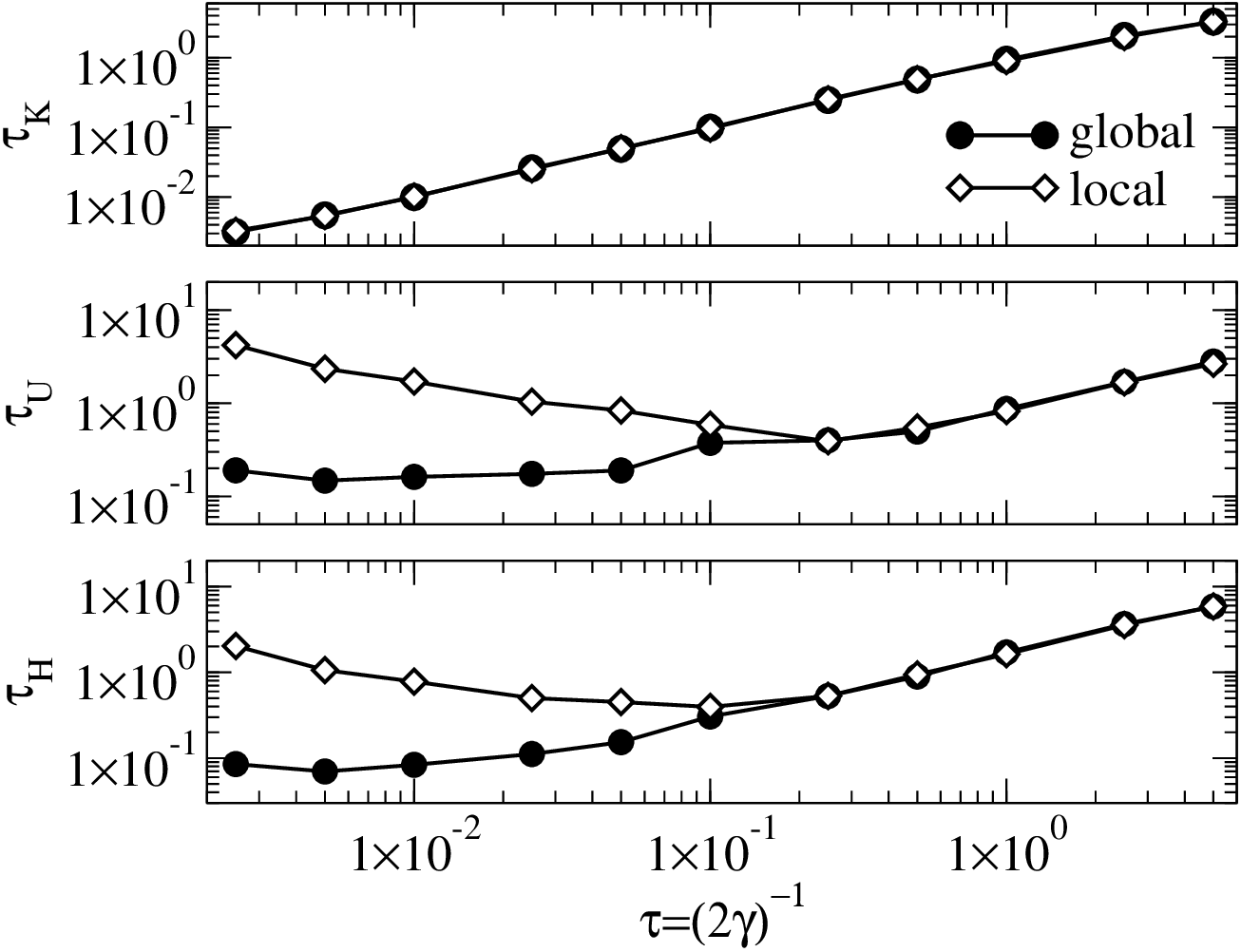}
\caption{
\label{fig-autocorrelation}
Autocorrelation time of the kinetic ($\tau_K$), potential ($\tau_U$),
and total ($\tau_H$) energy, as a function of the thermostat relaxation time $\tau$, 
for the local and global thermostats as indicated. The statistical error
is smaller than the symbol size.
All the quantities are in Lennard-Jones reduced units.
}
\end{figure}

In Figure~\ref{fig-autocorrelation} we plot the autocorrelation time
of the kinetic energy $\tau_K$, of the potential energy $\tau_U$
and of the total energy $\tau_H$, as a function of the
thermostat relaxation time $\tau$.
The autocorrelation time of the kinetic energy is
completely dictated by the thermostat relaxation time, and independent
on the choice of a local or a global scheme. On the contrary,
the autocorrelation time of the total energy and of the potential energies
are proportional to $\tau$ only in the limit of large $\tau$.
In the local scheme, when $\tau$ is smaller than 0.2 the disturbance
of the trajectory becomes so large that the phase-space exploration turns out to
be slower. Comparing Figs.~\ref{fig-diffusion}
and~\ref{fig-autocorrelation}, it is seen that the optimal value for
$\tau$ is the smallest one that still does not
affect the diffusion coefficient.
When the global scheme is adopted, even small values of $\tau$ can be safely
used, resulting in a faster decorrelation of the total energy.

\section{Conclusions}

In conclusion, we have presented an alternative derivation of the global
thermostat introduced in Ref.~\cite{buss+07jcp}. This derivation
allows to write continuous equations of motion and
shows the analogy between this scheme and the
standard Langevin thermostat. Namely, the new scheme can be considered
as a global version of the Langevin thermostat,
that minimizes the disturbance of the original Hamiltonian dynamics.
Finally, we have discussed these properties on a simple test case,
showing that the global scheme preserves the dynamical properties.
Moreover, using as a measure the autocorrelation time of the total
energy, we have shown that the global scheme allows for a faster
sampling.

\end{document}